\documentclass[11pt]{article}
\usepackage{amsmath,amssymb,bm,epsf,epsfig,graphicx,hyperref,physics, cleveref, subfig, longtable, array, multirow, ragged2e}
\usepackage[backend=bibtex,natbib,style=numeric-comp,sorting=none, doi=false, isbn=false,url=false]{biblatex}
\input epsf.sty
\newcommand{\sectiono}[1]{\section{#1}\setcounter{equation}{0}}

\newcommand{\lambdaB}[2]{\{#1\, _\lambda \,#2 \}}

\def\ket#1{|#1 \rangle}

\def \be {\begin{equation}}
\def \ee {\end{equation}}
\def \bea {\begin{eqnarray}}
\def \eea {\end{eqnarray}}
\def \bdm {\begin{displaymath}}
\def \edm {\end{displaymath}}

\def \ra {\rightarrow}
\def \pad {\partial_{\alpha}}

\def\0{\nonumber}

\begin{document}
\begin{center}
 {\large \bf $\,$\\
\vskip2cm
The semi-chiral ring of supersymmetric $\phi^4$ theory as a representation}

\vskip 1.1cm

{\large Jaroslav Scheinpflug\footnote{Email:
jscheinpflug@g.harvard.edu}$^{(a), (b)}$
} \vskip 1 cm

$^{(a)}${\it {Perimeter Institute for Theoretical Physics} \\
{Waterloo, ON N2L 2Y5, Canada}} \\
$^{(b)}${\it{Jefferson Physical Laboratory, Harvard University} \\ {Cambridge, MA 02138 USA}}
\end{center}

\vspace*{6.0ex}

\centerline{\bf Abstract}
In this short note, we study the infinite-dimensional symmetry algebras which appear in holomorphic twists of 4d $\mathcal{N}=1$ supersymmetric quantum field theories. In particular, we investigate whether their representation theory helps us understand the semi-chiral ring of $\frac{1}{4}$-BPS operators. We focus on the supersymmetric analogue of $\phi^4$ theory. Upon twist, this becomes a 4d $\beta \gamma$ system deformed by a cubic superpotential. We compute the semi-chiral ring in this example and organize it into modules for the algebra generated by the stress tensor. We find an intricate module structure and falsify the hypothesis that this could be a 4d analogue of the 2d Ising Virasoro minimal model.
\bigskip

\vfill \eject

\baselineskip=15pt

\tableofcontents

\section{Introduction and summary}

Supersymmetric quantum field theories (SQFTs) have access to quantities that are protected from receiving various quantum corrections. This has lead to major progress in understanding the dynamics of such theories, see \cite{Intriligator:1995au,Tachikawa:2018sae,Razamat:2022gpm} for reviews in the context of 4d $\mathcal{N}=1$ theories. Working in this context, we study a \textit{semi-chiral ring} of $\frac{1}{4}-$BPS local operators \cite{Budzik:2023xbr} i.e. of those protected by being annihilated by one of the four supercharges $Q$. The information contained in this ring is neatly captured by passing to the $Q$-cohomology subsector of the original theory i.e. by \textit{twisting} by $Q$ \cite{Witten:1988ze}, which in our setup yields a cohomological \textit{holomorphic theory} \cite{phdthesis, costello2013notes, Eager:2018oww} defined on $\mathbb{C}^2$, having two holomorphic directions. Such a \textit{holomorphic twist} possesses additional symmetries, their infinite-dimensional algebras \cite{Saberi:2019fkq} being analogous to 2d chiral algebras.  These algebras have recently been systematically studied in perturbation theory \cite{Budzik:2022mpd, Budzik:2023xbr, DavideNewPaper}, revealing that the various algebraic operations, including the action of $Q$, generically receive quantum corrections, leading to interesting effects such as holomorphic confinement \cite{Budzik:2023xbr}. Another interesting recent development in the subject is \cite{Bomans:2023mkd}, which shows that the holomorphic twist has the potential to teach us about the physical theory since there are higher brackets encoding the $a$ and $c$ anomaly coefficients.

In this note, we investigate whether the representation theory of these infinite-dimensional algebras can be used to elucidate the structure of the semi-chiral ring. We do this on a particularly simple example of the theory of a single free chiral multiplet deformed by a cubic superpotential (supersymmetric analogue of $\phi^4$), whose holomorphic twist is the 4d $\beta \gamma$ system deformed by $\gamma^3$ \cite{Saberi:2019ghy}. In particular we wish to test a scenario, where this theory would be analogous to a minimal model and the semi-chiral ring would consist of higher Virasoro \cite{Saberi:2019fkq, Bomans:2023mkd} descendants of ground states. To test this scenario, we compute the semi-chiral ring as the BRST cohomology of the holomorphic twist, try to identify possible ground states and see whether the whole ring is generated by acting on them with the modes of the stress tensor $S_\alpha = \frac{2}{3}\beta \partial_\alpha \gamma - \frac{1}{3} \gamma \partial_\alpha \beta $, $\alpha \in 1,2$. To be more specific, we act with the positive modes by computing the $\lambda$-2-bracket (which up to the weight we consider is tree-level exact) and the negative modes by computing the loop-corrected regularized product \cite{Budzik:2022mpd, Budzik:2023xbr, DavideNewPaper} with the $S_\alpha$. It is interesting to see whether the loop corrections to products play a role, possibly being needed to simplify the representation theory. Note that we do all these calculations using Mathematica to perform the symbolic manipulation. The code that computes the $\lambda$-2-bracket, loop corrected regularized product and the semi-chiral ring is attached \footnote{Also available at \href{https://github.com/jscheinpflug/semi-chiral-representation}{https://github.com/jscheinpflug/semi-chiral-representation}} (with some examples of usage). Note that although the semi-chiral ring is computed up to weight 10, its representation-theoretic structure is investigated only up to weight 6.

We find that the semi-chiral ring possesses an intricate structure and although a large part of it is indeed covered by acting with modes of $S_\alpha$ on three ground states $1, \gamma$ and $\xi_{\alpha \beta} \equiv \partial_\alpha \gamma \partial_\beta \gamma - 2 \gamma \partial_\alpha \partial_\beta \gamma$, there are representatives that cannot be written in that way starting from three ground states, for example the scalar $\Upsilon \equiv \frac{1}{4}\gamma \partial_\alpha \beta \partial^\alpha \beta - \partial_\alpha \gamma \beta \partial^\alpha \beta$ and the spin 2 tensor $\gamma \partial_\alpha \partial_\beta \partial_\gamma \partial_\delta \gamma$. Thus by ruling out a three-tower structure, these fields have obstructed the theory being a higher Virasoro analogue of the 2d Ising model. One may ask whether the loop corrections have helped in filling out the cohomology in any way, generating fields that were not obtained as higher Virasoro descendants. Up to the level we considered, they had only one such  effect and that is to lift $S_\alpha S^\alpha = \frac{1}{9} Q_B(\beta \partial_\alpha \beta \partial^\alpha \beta)$ into cohomology i.e. the loop-corrected product $(S_\alpha \, S^\alpha)$ is in cohomology. Although this by itself did not enable the simplest minimal model construction, a simple counting argument shows that such loop corrections are much more abundant at higher levels and so they might still play an important role in realizing a non-Ising-like scenario.

At this point, we can only speculate about what such a scenario might look like. We are naturally presented with two possibilities. One is that the theory is still a higher Virasoro minimal model, but it is not the simplest one, not having three towers like the 2d Ising. A realization of this that is consistent with our data is if the two cohomology elements $\Upsilon$ and $\gamma \pad \partial_\beta \partial_\gamma \partial_\delta \gamma$ are added as ground states, although it seems unlikely that the number of ground states would truncate at five. The other possibility, which we thus find more natural, is that the theory is a minimal model of an extension of the higher Virasoro algebra. An example of such an extension that is consistent with our data is to add the charged $\gamma$ as an additional generator to our algebra and take as ground states $1, \Upsilon$. A more intrinsic way of enhancing the algebra would be to consider various higher products that are known to be nontrivial \cite{Bomans:2023mkd}.

Due to the semi-chiral ring being rich in structure, it seems that in the future, it would be valuable to adapt a bottom-up approach in addition to the top-down approach used here. After understanding the kinematics in such an approach, one might also try to tackle dynamical questions and perhaps set up a bootstrap by searching for null vectors. This would constrain certain combinations of brackets to be zero, which is analogous to the usual hypergeometric differential equations of 2d CFT \cite{Belavin:1984vu}. This direction presents a major challenge for the future.

The note is organized as follows. In section (\ref{setupSec}), we review the setup by defining the theory of interest and the algebraic operations in it. In section (\ref{resultsSec}), we show how we calculate the semi-chiral ring and see how it fits into representations. Appendix \ref{semi-chiral ring app} explicitly lists the elements of the semi-chiral ring up to weight 6 and appendix \ref{partFapp} summarizes the semi-chiral ring up to weight 10 by using it to compute the partition function and the supersymmetric index.

\section{Setting up the holomorphic twist}
\label{setupSec}
In this section, we set up the specific holomorphic twist of interest and briefly review the algebraic structures, whose representations we investigate in the next section. We largely follow the works of \cite{Saberi:2019ghy}, who showed how the theory of a deformed chiral multiplet becomes the deformed 4d $\beta \gamma$ system upon twist and of \cite{Budzik:2022mpd, Budzik:2023xbr, DavideNewPaper}, who studied the algebras of the $\mathcal{N}=1$ holomorphic twist in perturbation theory. Part of the discussion is also inspired by the introduction of \cite{Bomans:2023mkd}.

\subsection{The holomorphic twist}
We consider our theory as defined on Euclidean $\mathbb{R}^4$. The supercharges are the two-component spinors $Q_{\dot{\alpha}}, \bar{Q}_{\alpha}$ (the dotted-undotted convention swapped since we twist soon). Here $\alpha \in \pm$, which we often write as $\alpha \in 1,2$. We choose one of the four supercharges $Q \equiv Q_{\dot{-}}$ to be the twisting supercharge to whose cohomology we pass, which using the usual $\mathcal{N}=1$ SUSY algebra
\be
\acomm{Q}{\bar{Q}_\alpha} = \partial_{\dot{-}\alpha} = \partial_{\bar{z}^\alpha}
\label{exactDel}
\ee
equips $\mathbb{R}^4$ with a complex structure such that the antiholomorphic direction $\bar{z}^\alpha = x^{\dot{-} \alpha}$ cohomologically drops out. One is then left only with two holomorphic directions $z^{\alpha} = x^{\dot{+} \alpha}$, making the theory defined on $\mathbb{C}^2$. This choice of supercharge breaks the $\text{Spin}(4) \simeq SU(2) \cross SU(2)$ rotations to an $SU(2)$ subgroup that leaves the holomorphic volume $\dd z^1 \wedge \dd z^2$ invariant. It is the spin under this subgroup that we will often classify our fields under.

The authors of \cite{Saberi:2019ghy} showed that upon twist the theory of a chiral $\Phi$ deformed by $W(\Phi)$ becomes the 4d $\beta \gamma$ system deformed by $W(\gamma)$ with action
\be
S_{\beta \gamma} = \int \dd^2 z \big(\beta \bar{\partial}\gamma + W(\gamma)\big)
\ee
where $\beta, \gamma$ are scalars ($\beta$ a fermion, $\gamma$ a boson).
Since the undeformed theory is free, one can evaluate all perturbative quantities of interest using the fact that the propagator is the \textit{Bochner-Martinelli kernel}
\be
P(z,\bar{z},\dd \bar{z}) = \frac{1}{4\pi^2}\frac{\bar{z}^2\dd \bar{z}^1 - \bar{z}^1 \dd \bar{z}^2}{\abs{z}^4} \equiv \omega_{BM},
\label{Bochner}
\ee
satisfying $\bar{\partial} P(z, \bar{z}, \dd\bar{z}) = \dd \bar{z}^1 \dd \bar{z}^2 \delta^4(z)$. Note that $P^2 = 0$, leading to dramatic simplifications in the structure of Feynman diagrams \cite{Budzik:2022mpd}.

In the specific twist we consider (twist of a supersymmetric analogue of $\phi^4$ theory), one has the superpotential $W(\gamma) = \frac{\gamma^3}{3}$. Classical conformal symmetry requires that we have $\Delta_\gamma = \frac{2}{3}$ and $\Delta_\beta = \frac{4}{3}$. Importantly, the potential preserves a $\mathbb{Z}_3$ global symmetry such that $\gamma \to z \gamma, \beta \to z^2 \beta$ with $z^3 = 1$, decomposing the spectrum into three towers. The twist has the weight 3 stress tensor
\be
S_\alpha = \frac{2}{3}\beta \partial_\alpha \gamma - \frac{1}{3} \gamma \partial_\alpha \beta,
\ee which is neutral so that acting with it does not move one across towers.

\subsection{Algebras of the holomorphic twist}
We continue by reviewing the algebraic structure of the $\mathcal{N}=1$ holomorphic twists and illustrate it on the twist described above.

We begin by defining the \textit{semi-chiral ring} \cite{Budzik:2023xbr}. It will prove useful to assemble our fields into \textit{superfields} defined on \textit{superspace} i.e. the base manifold equipped with extra Grassmann odd coordinates $\theta^{\dot{\alpha}}, \bar{\theta}^\alpha$. The superfield $\mathcal{O}[\bar{\theta}] = e^{\bar{\theta} \bar{Q}} \mathcal{O} = \mathcal{O}^{(0)} + \mathcal{O}^{(1)} + \mathcal{O}^{(2)}$ can be thought of as a Dolbeault form upon $\bar{\theta}^\alpha \leftrightarrow \dd \bar{z}^\alpha$, where the Dolbeault operator is $\bar{\partial} = \bar{\theta}^\alpha \partial_{\bar{z}^\alpha}$. Thinking of fields as being assembled into forms like this is useful since it simplifies some of the SUSY algebra and lets us naturally integrate over expressions made out of fields (the algebraic operations being defined as integrals). If the superfield is annihilated by the superderivative: $D_{\dot{-}} \mathcal{O} = (Q + \bar{\partial}) \mathcal{O} = 0$, we call it a \textit{semi-chiral superfield}. The bottom component $\mathcal{O}^{(0)}$ is then a \textit{semi-chiral operator}, being annihilated by $Q$. From the twist being cohomologically holomorphic, it follows that the OPE of two semi-chiral operators is meromorphic. Since there are no meromorphic functions with isolated singularities on $\mathbb{C}^2$ (Hartogs's extension theorem), we conclude that the OPE of two semi-chiral operators is regular and so they form a ring, the semi-chiral ring. This gives one more reason to consider entire superfields since that gives the holomorphic twist a direct access to singularities in the OPE. Of course, all of this information is already hidden in the semi-chiral ring, the $Q$-cohomology being isomorphic to the space of superfields modulo the action of $Q + \bar{\partial}$ \cite{Budzik:2023xbr}. To give an example, in the free 4d $\beta \gamma$ system, both $\beta$ and $\gamma$ are semi-chiral superfields. Upon deforming by $W(\gamma)$, the superfield $\beta$ is no longer semi-chiral as we show below.

We now define various algebraic operations on semi-chiral superfields \cite{Budzik:2022mpd, Budzik:2023xbr, Bomans:2023mkd, DavideNewPaper}. Just as in the 2d chiral algebra context one can encode the OPE via the products $\acomm{\bullet}{\bullet}_n$, which arise by writing
\be
\mathcal{O}_1(z) \mathcal{O}_2(0) = \sum\limits_{n \in \mathbb{Z}} \frac{\acomm{\mathcal{O}_1}{\mathcal{O}_2}_n}{z^{n+1}} \leftrightarrow \acomm{\mathcal{O}_1}{\mathcal{O}_2}_n = \oint\limits_{S^1} \frac{\dd z}{2 \pi i} z^n \mathcal{O}_1(z)\mathcal{O}_2(0),
\ee
one can do a similar construction in our context. This time, the modes are in correspondence with the Dolbeault cohomology elements $H^{0, \bullet}(\mathbb{C}^2 \backslash\{0\})$, which are concentrated in degrees $(0,0)$ and $(0,1)$, the former being polynomials $\mathbb{C}(z_1, z_2)$ and the latter being holomorphic derivatives of the Bochner-Martinelli kernel $\mathbb{C}(\partial_1,\partial_2) \, \omega_{BM}$. Thus, we have the non-negative modes
\be
\acomm{\mathcal{O}_1}{\mathcal{O}_2}_{n_1, n_2} = \oint\limits_{S^3} \frac{\dd^2 z}{(2\pi i)^2} z_1^{n_1} z_2^{n_2} \mathcal{O}_1(z) \mathcal{O}_2(0)
\ee
with $n_1, n_2 \geq 0$ and the negative modes
\be
\acomm{\mathcal{O}_1}{\mathcal{O}_2}_{n_1, n_2} = \oint\limits_{S^3} \frac{\dd^2 z}{(2\pi i)^2} \partial_1^{-n_1} \partial_2^{-n_2} \omega_{BM} \mathcal{O}_1(z) \mathcal{O}_2(0)
\ee
with $n_1, n_2 \leq -1$. One can capture the non-negative modes using the \textit{$\lambda$-2-bracket}
\be
\lambdaB{\mathcal{O}_1}{\mathcal{O}_2} = \oint\limits_{S^3}\frac{\dd^2 z}{(2\pi i)^2} e^{\lambda \cdot z} \mathcal{O}_1(z) \mathcal{O}_2(0) = Q\int\limits_{\mathbb{C}^2} \frac{\dd^2 z}{(2\pi i)^2} e^{\lambda \cdot z} :\mathcal{O}_1(z)\mathcal{O}_2(0):
\label{lambdabracket}
\ee
where in the second expression valid in free field theory, we used that $Q$ turns into a $\bar{\partial}$ when acting on semi-chiral superfields. Analogously, in free theories, one can define higher $\lambda$-brackets such as the \textit{$\lambda$-3-bracket }
\be
\{\mathcal{O}_1 \,_{\lambda_1} \mathcal{O}_2 \,_{\lambda_2} \mathcal{O}_3 \} \equiv Q\hspace{-0.2cm}\int\limits_{(\mathbb{C}^2)^2} \frac{\dd^2 z_1 \dd^2 z_2}{(2\pi i)^4} e^{\lambda_1\cdot z_1 + \lambda_2 \cdot z_2} :\mathcal{O}_1(z_1)\mathcal{O}_2(z_2) \mathcal{O}_3(0):
\label{lambdaThree}
\ee
The negative modes are captured by the \textit{regularized product}
\be
( \mathcal{O}_1 \,_\lambda\, \mathcal{O}_2 ) = \oint\limits_{S^3} \frac{\dd z^2}{(2\pi i)^2} e^{\lambda \cdot z} P(z) \mathcal{O}_1(z) \mathcal{O}_2(0),
\ee
which is essentially the $\lambda$-2-bracket, but with an extra propagator put in by hand. This will nicely reflect in the diagrammatic rules that we present below.

Since we are interested in an interacting theory, it is likely that deforming the products appropriately might make them more useful to for example organize the interacting spectrum into representations. The definition of the deformed brackets is rather intuitive, following the usual picture of inserting an exponential of an interaction $\mathcal{I}$, giving us (where there is no $\lambda$, it is assumed  to be zero)
\be
\begin{aligned}
\{\mathcal{O}_1 \, _\lambda \, \mathcal{O}_2 \}_{\mathcal{I}} &\equiv \{ \mathcal{O}_1 \, _\lambda \, \mathcal{O}_2 \} + \{\mathcal{I} \, \mathcal{O}_1 \, _\lambda \, \mathcal{O}_2 \} + \frac{1}{2!} \{ \mathcal{I} \,\mathcal{I} \,\mathcal{O}_1 \, _\lambda \,\mathcal{O}_2 \} + \ldots \\
    ( \mathcal{O}_1 \, _\lambda \, \mathcal{O}_2 )_{\mathcal{I}} &\equiv ( \mathcal{O}_1 \, _\lambda \, \mathcal{O}_2 ) + (\mathcal{I} \, \mathcal{O}_1 \, _\lambda \, \mathcal{O}_2 ) + \frac{1}{2!} ( \mathcal{I} \,\mathcal{I} \,\mathcal{O}_1 \, _\lambda \,\mathcal{O}_2 ) + \ldots
    \label{deformLabmda}
\end{aligned}
\ee
and their higher analogues. To see that this intuitive definition is indeed appropriate, follow \cite{DavideNewPaper} who show that there is a scheme in which the $\lambda$-brackets capture the BRST anomaly of a deformed holomorphic twist as follows:
 \be
 \boldsymbol{Q_B} \mathcal{O} = Q_B \mathcal{O} +  \{\mathcal{I} \,\mathcal{O}\} +\frac{1}{2!} \{\mathcal{I} \,\mathcal{I} \,\mathcal{O}\} +\ldots,
 \label{deformedQ}
 \ee
 where $\boldsymbol{Q_B}$ is the deformed $Q_B$. In other words, the $\lambda$-brackets capture the $L_\infty$ structure \cite{Zwiebach:1992ie, Stasheff:1993ny, Hohm:2017pnh, Doubek:2020rbg} of the deformed theory. Then upon writing $\mathcal{I} \to \mathcal{I} + \mathcal{I}'$ and reorganizing the new expansion in terms of $\mathcal{I}'$-deformed brackets, one gets that (\ref{deformLabmda}) follows (for the $\lambda$-bracket case). The interaction preserves BRST symmetry if the \textit{Maurer-Cartan equation} $\boldsymbol{Q_B} \mathcal{I} = 0$ is satisfied.

 We want to illustrate the above on the deformed 4d $\beta \gamma$ system and to do this, the Feynman diagrammatics of \cite{Budzik:2022mpd} come in very handy. The vertices of the Feynman diagrams correspond to the entries of brackets and the edges correspond to propagators. For example at tree level, diagrammatically, the $\lambda$-2-bracket and the regularized product are represented as follows \\
 \begin{figure}[h!]
	    \centering
    \subfloat[\centering Tree-level $\lambda$-2-bracket]{{\includegraphics[width=8cm]{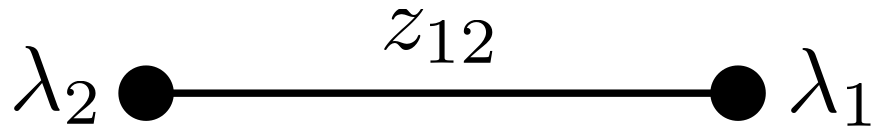} }}%
    \quad
    \subfloat[\centering Tree-level regularized product]{{\includegraphics[width=8cm]{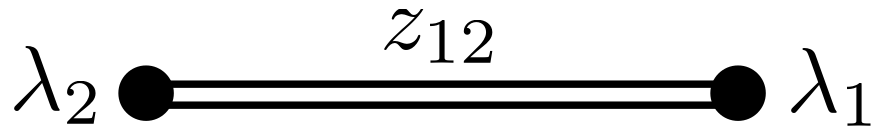} }}%
    \caption{Tree level $\lambda$-2-bracket and the regularized product. Single line represents a propagator coming from a nontrivial contraction and a double line represents a propagator inserted by hand.}
	    \label{vertex}
\end{figure}
\\

It was shown in \cite{Budzik:2022mpd} that the Feynman diagrams one obtains are heavily restricted due to $\omega_{BM}^2 = 0$ and correspond to Laman graphs \cite{Laman1970OnGA}, which are \textit{minimally rigid} i.e. they are rigid (only degrees of freedom being planar rotations and translations) unless an edge is removed. Up to two loops, the only Laman graphs are the segment, the triangle and the bitriangle giving e.g. the following perturbative expansion of the deformed $\lambda$-2-bracket\\
  \begin{figure}[h!]
	    \centering
\includegraphics[width=17cm]{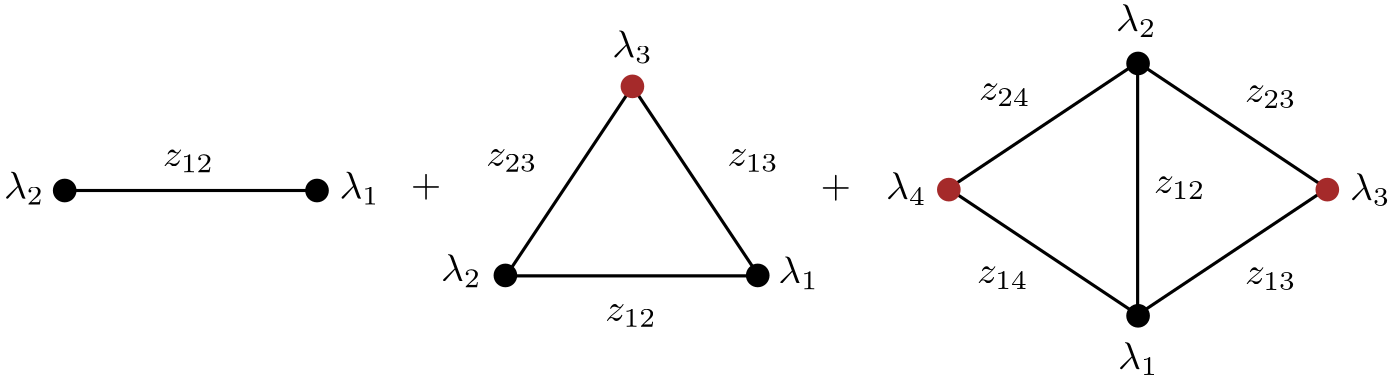}
    \caption{Contributions to the $\lambda$-2-bracket up to two loops. The interaction vertices are colored in brown.}
	    \label{vertex}
\end{figure}
\\
 The key idea here is that one can associate to Feynman diagrams theory-independent Feynman integrals that are calculated once and for all and appropriately dress them by theory-specific factors to get the answers for either the $\lambda$-brackets or the regularized product. Since in our theory of interest the propagator is strictly the Bochner-Martinelli kernel, we will not need to dress the Feynman integrals. Note that if we wish to compute the brackets containing a derivative of a field, we simply perform a rigid shift of its position, Taylor expand and at the end of the calculation collect the appropriate coefficient.
 The Feynman integrals are polynomials in the various momenta $\lambda_
 \alpha$ and rigid shifts of operators $z^\alpha$ and when we encounter an uncontracted operator $\mathcal{O}(w)$ with $w$ integrated over, we write it as $e^{w \cdot \partial} \mathcal{O}(0)$ so that having it effectively corresponds to shifting a given momentum $\lambda_\alpha \to \lambda_\alpha + \partial_\alpha$. The Feynman integrals for the segment and the triangle graph are \cite{Budzik:2022mpd}
 \begin{align}
\mathcal{I}_{seg.}(\lambda, z) &= e^{-\lambda_1 \cdot z_{12}} = e^{\lambda_2 \cdot z_{12}} \\
\label{Is}
\mathcal{I}_{triang.}(\lambda, z) &= -e^{\lambda_2 \cdot z_{12} + \lambda_3 \cdot z_{13}}\lambda_2 \wedge \lambda_3
\biggr[\frac{e^{-\lambda_2 \cdot Z}}{(\lambda_1 \cdot Z)(\lambda_2 \cdot Z)} + \frac{e^{\lambda_3 \cdot Z}}{(\lambda_1 \cdot Z)(\lambda_3\cdot Z)} + \frac{1}{(\lambda_2 \cdot Z)(\lambda_3 \cdot Z)}
\biggr],
\end{align}
where in these expressions an extra $\lambda$ is added by hand so that translation invariance $\sum\limits_i \lambda_i = 0$ holds and $\lambda_2 \wedge \lambda_3 \equiv {\lambda_2}_+ {\lambda_3}_- - {\lambda_2}_- {\lambda_3}_+$.

 We think that this formalism is best illustrated by examples and so let us display a few. To compute the BRST charge of the 4d $\beta \gamma$ system deformed by $\mathcal{I} = \frac{\gamma^3}{3}$, we can use the fact that only $\beta$ and $\gamma$ have a nonzero contraction. This means that the only relevant Feynman diagram is the segment since it is impossible to contract the fields to make other diagrams and $\boldsymbol{Q_B}$ is tree-level exact. Thus
 \be
 \begin{aligned}
 \boldsymbol{Q_B} \gamma &= Q_B \gamma = 0 \\
  \boldsymbol{Q_B} \beta &= Q_B \beta + \{\frac{\gamma^3}{3} \, \beta \} = \gamma^2
  \label{QAct}
 \end{aligned}
 \ee
 where we used that $Q_B \gamma = Q_B \beta = 0$ in the free theory and that $\{ \frac{\gamma^3}{3} \, \beta\} = \mathcal{I}_{seg.}(0 + \partial, 0) \gamma^2 = \gamma^2$ (the $\frac{1}{3}$ drops out since there are three possible Wick contractions). More generally $\boldsymbol{Q_B} \gamma = 0, \boldsymbol{Q_B}  \beta = \partial W(\gamma)$. Note that these deformations leads to a consistent theory since the Maurer-Cartan equation $\boldsymbol{Q_B} W = 0$ is trivially satisfied. One could also consider the marginal deformation by the ghost number current $J = \beta \gamma$ and then $\boldsymbol{Q_B} \gamma = \gamma, \boldsymbol{Q_B} \beta = \beta$ and $\boldsymbol{Q_B} J= 0$, again leading to a consistent theory.

\begin{figure}[h!]
	\centering
    {\includegraphics[width=8cm]{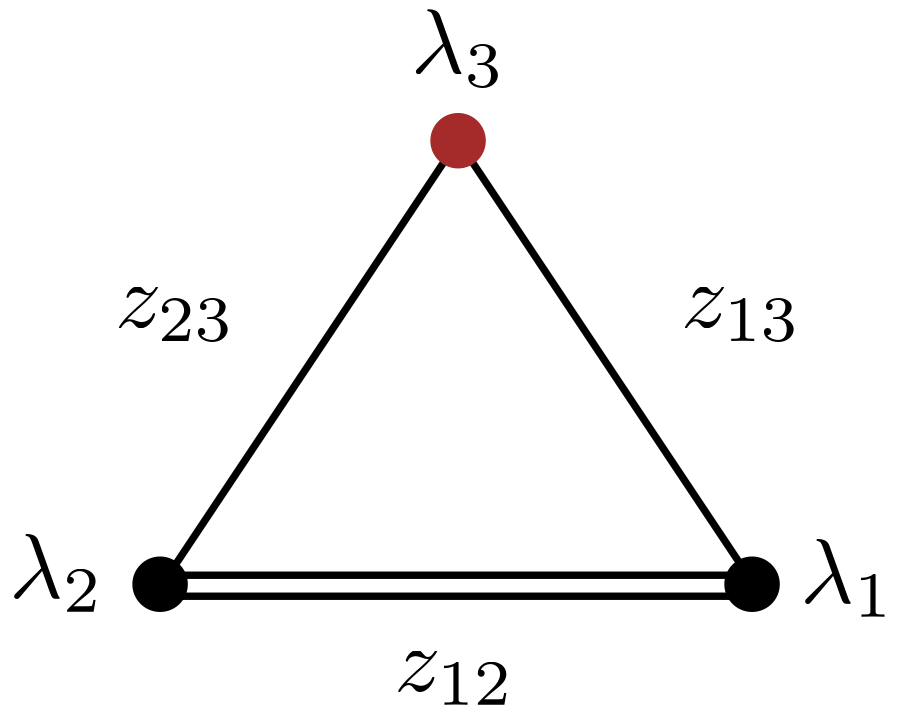} }%
    \caption{Loop correction to the regularized product. The two stress tensors are joined by a rigid propagator and nontrivially contract with the interaction.}
	    \label{vertex2}
\end{figure}

We now give an example of a nontrivial loop correction to the regularized product. The product we consider is $(S_\alpha \, S^\alpha)$, with the index raised by the antisymmetric $\epsilon^{\alpha \beta}$. It gets a loop correction since there are two contractions with $\beta$ fields and $\gamma^3$ that combine with the rigid $\omega_{BM}$ factor of the regularized product to form the three sides of a triangle, see figure (\ref{vertex2}).
\\
Using (\ref{Is}), we calculate it in Mathematica and obtain
\be
(S_\alpha \, S^\alpha) = S_\alpha S^\alpha + \frac{10}{27}\gamma(\partial_1\partial_2\gamma)^2-\frac{16}{27}\partial_2^2\gamma(\partial_1\gamma)^2-\frac{16}{27}\partial_1^2\gamma(\partial_2\gamma)^2-\frac{10}{27}\gamma\partial_2^2\gamma\partial_1^2\gamma+\frac{32}{27}\partial_2\gamma\partial_1\gamma\partial_1\partial_2\gamma.
\label{sLoop}
\ee

\section{Results}
\label{resultsSec}
In this section, we show how we compute the semi-chiral ring as the BRST cohomology of the deformed 4d $\beta \gamma$ system and then attempt to organize it into representations.
We drop the boldface on $\boldsymbol{Q_B}$ in the following.

\subsection{Computing the semi-chiral ring}
The semi-chiral ring is computed as the BRST cohomology of the BRST charge defined by (\ref{QAct}).
Since $Q_B$ is a linear operator, computing its cohomology is a simple linear algebra problem.

We use a trick in that we compute the cohomology as ground states of the "Hamiltonian" \cite{Witten:1982im} \footnote{Note that this Hamiltonian need not necessarily be a physical Hamiltonian of our theory}
\be
H \equiv \acomm{Q_B}{Q_B^\dagger},
\ee
where $Q_B^\dagger$ is Hermitean-conjugated with respect to an inner product that we specify momentarily. The idea behind this is that $H$ is a positive operator, which gives that
\be
H \ket{\Psi} = 0 \leftrightarrow Q_B\ket{\Psi} = Q_B^\dagger \ket{\Psi} = 0.
\ee
This clearly means that the ground states of $H$ are all $Q_B$-closed. Are there $Q_B$-closed states that are not ground states? That is, can there be states in the cohomology such that $Q_B\ket{\Psi} = 0$, but $Q_B^\dagger \ket{\Psi} \neq 0$? If there was such a state, then we could write for it
\be
\ket{\Psi} = \acomm{Q_B}{Q_B^\dagger}\frac{1}{H} \ket{\Psi}
\ee
so it wouldn't be in the cohomology because it would be a sum of a $Q_B$-exact state and a state that is not $Q_B$-closed. The statement here can be rephrased by saying that states which are not ground states cancel pairwise in $Q_B$-cohomology (the $Q_B, Q_B^\dagger$ form a two-dimensional Clifford algebra and so the states can be projected onto a $Q_B$-exact and a non-$Q_B$-closed part with the $Q_B, Q_B^\dagger$ acting as ladder operators between them). Lastly, one should also ask whether all the exact states decouple. Since $\comm{Q_B}{H}=0$, this means that the $Q_B-$exact state would have to be a $Q_B$-image of a ground state. But since ground states are $Q_B$-closed, the state would be zero and so exact states indeed decouple.

Now we fill in the details on our choice of inner product. We wish to choose our inner product so that $H$ preserves the symmetries we want. In particular, we want it to preserve the $SU(2)$ rotations unbroken by the choice of the twisting supercharge. If we can find such an inner product, then we can simplify our cohomology computation by restricting to the subspace of highest weight states of the $SU(2)$. The entire cohomology is then generated from the highest weight cohomology by acting on it with lowering operators. One can represent $SU(2)$ on fields regarded as two-variable monomials in derivatives. The action of raising and lowering operators in this representation is as follows
\be
\begin{aligned}
    J_+ \partial_1^n \partial_2^m \gamma &= m \partial_1^{n+1} \partial_2^{m-1} \gamma \\
        J_- \partial_1^n \partial_2^m \gamma &= n \partial_1^{n-1} \partial_2^{m+1} \gamma \\
    J_+ \partial_1^n \gamma &= 0 \\
    J_- \partial_2^m \gamma &= 0
\end{aligned}
\ee
and similarly on derivatives of $\beta$. Action on other operators follows from the Leibniz rule and linearity. We can also form the Cartan $J_3 = \frac{1}{2}\comm{J_+}{J_-}$ that counts the number of 1-derivatives minus the number of 2-derivatives. We tested in Mathematica that $\comm{Q_B}{J_\pm} = 0$ and so for $H$ to commute with the $SU(2)$ generators, it is enough to find an inner product such that $(J_+)^\dagger = J_-$. To find this inner product, we can view our system as having two ground states $\beta, \gamma$ and the $\partial_1$ and $\partial_2$ are acting like bosonic creation operators on it. Thus we can try to introduce the usual inner product on such a Fock space:
\begin{equation}
    \begin{aligned}
        \braket{\partial_1^n \partial_2^m \gamma}{\partial_1^n \partial_2^m \gamma} =  \braket{\partial_1^n \partial_2^m \beta}{\partial_1^n \partial_2^m \beta} = \frac{n! m!}{(n+m)!},
    \end{aligned}
\end{equation}
from which follows e.g. the useful $\braket{(\partial_1^n \partial_2^m \gamma)^k}{(\partial_1^n \partial_2^m \gamma)^k} = k! (\frac{n! m!}{(n+m)!})^k$. The inner product vanishes otherwise. Since this inner product is not orthonormal, we Hermitean conjugate as $A^\dagger = G^{-1} (A^*)^T G$, where $G$ is the Gram matrix. We can now try to show that indeed $(J_+)^\dagger = J_-$ on single fields by a direct calculation:
\begin{equation}
    \begin{aligned}
        \braket{\partial_1^n \partial_2^m \gamma}{J_+\partial_1^{n-1} \partial_2^{m+1} \gamma} &= (m+1) \frac{n! m!}{(n+m)!}\\
        \braket{J_- \partial_1^n \partial_2^m \gamma}{\partial_1^{n-1} \partial_2^{m+1} \gamma} &= n \frac{(n-1)! (m+1)!}{(n+m)!}\\
        \braket{\partial_1^{n+1} \partial_2^{m-1} \gamma(\partial_1^{n} \partial_2^{m} \gamma)^{k-1}}{J_+(\partial_1^{n} \partial_2^{m} \gamma)^k} &= k \frac{(n+1)! m!}{(n+m)!} (k-1)! \biggr(\frac{n! m!}{(n+m)!}\biggr)^{k-1} \\
       \braket{J_-(\partial_1^{n+1} \partial_2^{m-1} \gamma(\partial_1^{n} \partial_2^{m} \gamma)^{k-1})}{(\partial_1^{n} \partial_2^{m} \gamma)^k} &= (n+1) k! \biggr(\frac{n ! m!}{(n+m)!}\biggr)^k
    \end{aligned}
\end{equation}
and similarly for $\beta$, where the first two and last two equations are pairwise equal as they should.
We also double checked in Mathematica that the inner product indeed gives $(J_+)^\dagger = J_-$ even on products of fields and so we can compute our cohomology in the space of highest weight states.
We note that the cohomology will be naturally graded by the conformal weight and the $SU(2)$ spin.\clearpage
Our algorithm can thus be summarized as follows:
\begin{enumerate}
    \item Find all the monomials in $\beta, \gamma$ and their derivatives at a given conformal weight $\Delta$
    \item Compute the kernel of $J_+$ on these monomials to find the highest weight states of $SU(2)$
    \item On the highest weight states, compute the matrix representations of $Q_B$ and $G$
    \item Use these to form the Hamiltonian $H = \acomm{Q_B}{Q_B^\dagger}$
    \item Compute the cohomology as the kernel of $H$
    \item Apply $SU(2)$ lowering operators on the highest weight cohomology to get the full cohomology
\end{enumerate}
We implemented this algorithm in Mathematica and we list the obtained cohomology in appendix \ref{semi-chiral ring app}. Note that although the cohomology is listed up to weight 6, our program can compute cohomology classes up to higher weight (we went up to 10), but for those we have not performed the representation-theoretic analysis that follows in the next subsection so we do not explicitly list them. The classes from weight 6 up to weight 10 were used only to compute the partition function and the index in appendix \ref{partFapp}. The index matches its expected analytic form. We proceed by attempting to fit the semi-chiral ring into a module structure.

\subsection{Representation theory of the semi-chiral ring}
We make a hypothesis that because of the global $\mathbb{Z}_3$ of the model ($\gamma$ has charge 1, $\beta$ charge 2) and it being a supersymmetric analogue of $\phi^4$, the semi-chiral ring will decompose into three towers that are generated by acting with the stress-tensor $S_\alpha = \frac{2}{3}\beta \partial_\alpha \gamma - \frac{1}{3} \gamma \partial_\alpha \beta$, either via the $\lambda$-2-bracket or the regularized product (possibly loop corrected). The various modes of this action form the higher Virasoro algebra \cite{Saberi:2019fkq, Bomans:2023mkd} and so we are making a hypothesis that the semi-chiral ring of our theory is one of its 2d Ising-like modules.

We must first find suitable ground states on top of which to build the towers. This means that we look for the fields of a given $\mathbb{Z}_3$ quantum number that have the smallest conformal weight. These are found to be $1, \gamma$ and $\xi_{\alpha \beta} = \pad \gamma \partial_\beta \gamma - 2 \gamma \pad \partial_\beta \gamma$.
Since the $\lambda$-2-bracket with $S_\alpha$ acts like a derivative, we can build other semi-chiral ring elements such as $\partial_{\alpha_1} \ldots \partial_{\alpha_n} \gamma$, $\pad \xi_{\beta \gamma}$ or $\pad \partial_\beta \xi^{\alpha \beta}$ that we found in \ref{semi-chiral ring app}. Finding them is a necessity since $\comm{Q}{\partial_\alpha} = 0$.

Acting with the regularized product is more nontrivial. Looking at the identity module, together with taking derivatives it generates the semi-chiral ring elements such as $S_\alpha$, $\pad S_\beta$ or $\pad S^\alpha$. We also find that the scalar field $-12\gamma(\partial_1\partial_2\gamma)^2-\partial_2^2\gamma(\partial_1\gamma)^2-\partial_1^2\gamma(\partial_2\gamma)^2+12\gamma\partial_2^2\gamma\partial_1^2\gamma+2\partial_2\gamma\partial_1\gamma\partial_1\partial_2\gamma$ found at weight 6 is in the same class (differs by an exact operator) as the loop-corrected product $(S_\alpha \, S^\alpha)$ computed in (\ref{sLoop}). Note that this is a particularly interesting example of a loop correction to products, since at tree level one has that $S_\alpha S^\alpha = \frac{1}{9} Q(\beta \partial_\alpha \beta \partial^\alpha \beta)$ and so the loop correction lifts it into cohomology. The $\gamma$ module also contains nontrivial semi-chiral ring elements such as $6 \pad \gamma S_\beta - 8\gamma \pad S_\beta$ and its derivatives (note that its not symmetric in $\alpha$ and $\beta$ so there are more nontrivial ways to take the derivative). Lastly, for the $\xi_{\alpha \beta}$ module we found no nontrivial semi-chiral ring elements up to weight 6 that come from the regularized product. One would expect that perhaps at weight $\frac{19}{3}$, the descendant $S^\alpha \xi_{\alpha \beta}$ would be in the semi-chiral ring, but this is not the case since $S^\alpha \xi_{\alpha \beta} = \frac{1}{6}Q(\partial_\alpha \partial_\beta \gamma \beta \partial^\alpha \beta - \partial_\beta \gamma \partial_\alpha \beta \partial^\alpha \beta)$.

Were the $1, \gamma, \xi_{\alpha \beta}$ and their descendants listed above enough to cover the whole semi-chiral ring of \ref{semi-chiral ring app} up to weight 6? Despite the loop corrections providing some help, the answer turns out to be negative. This is because there are at least two fields that do not fit into the proposed modules. These are the weight $\frac{16}{3}$ scalar
\be
\Upsilon = \frac{1}{4} \gamma \partial_\alpha \beta \partial^\alpha \beta - \partial_\alpha \gamma \beta \partial^\alpha \beta
\ee
and the spin 2 tensor $\gamma \pad \partial_\beta \partial_\gamma \partial_\delta \gamma$ of the same weight. Due to their $\mathbb{Z}_3$ charge, both would have to be in the $\xi_{\alpha \beta}$ tower, but it is clear that e.g. $\Upsilon$ is not there since it has 2 $\beta$ fields so it would have to come from two actions of $S_\alpha$ on $\xi_{\alpha \beta}$, but $2\cdot 3 + \frac{10}{3} > \frac{16}{3}$. We thus falsify the hypothesis presented above and we have to seek its alternative.

A possibility that we think is very natural is that if one hopes to obtain a minimal model, one has to extend the higher Virasoro algebra. This can be done in a multitude of ways. One way would be to add to $S_\alpha$ a generator $\gamma$ that generates $\beta$ translations. Our data up to weight 6 would then be consistent with $S_\alpha$ and $\gamma$ acting on the two ground states $1$ and $\Upsilon$. Another, more intrinsic, way of enhancing the algebra would be to add higher brackets that are known to be nontrivial \cite{Bomans:2023mkd}. Note that these algebras may have to be loop-corrected as we saw a hint of the loop corrections helping to fill in the semi-chiral ring. Perhaps a less natural way the semi-chiral ring would fit into representations is that one would add $\gamma \pad \partial_\beta \partial_\gamma \partial_\delta \gamma$, $\Upsilon$ and possibly some other fields to the ground states we considered. It is not clear whether this procedure would terminate at a finite number of ground states, the simple theory perhaps not being a higher Virasoro minimal model at all. We leave further exploration of these possibilities to the future.

\section*{Acknowledgments}

We are grateful to Davide Gaiotto for suggesting the project and many helpful conversations and comments on the draft. We also thank Haitian Xie for comments on the draft. The author was supported by the Perimeter Institute for Theoretical Physics. Research at Perimeter Institute is supported in part by
the Government of Canada through the Department of Innovation, Science and Economic
Development Canada and by the Province of Ontario through the Ministry of Colleges and
Universities.

\appendix

\clearpage

\sectiono{The semi-chiral ring}
\label{semi-chiral ring app}

In this appendix, we list members of the semi-chiral ring of the four-dimensional $\beta \gamma$ system deformed by a cubic superpotential up to weight 6. For each highest weight representative, we show how it gets lifted to the full $SU(2)$ multiplet e.g. $\partial_1 \gamma \ra \partial_\alpha \gamma$. When the lifted fields are separated by a comma, it indicates taking their linear combination.

\renewcommand{\arraystretch}{1.8} 

\begin{longtable}{|>{\centering\arraybackslash}p{1.5cm}|>{\centering\arraybackslash}p{1.5cm}|p{12.5cm}|} 
\hline
\textbf{Weight} & \textbf{Spin} & \textbf{Semi-chiral ring element} \\
\hline
\endfirsthead 
\hline
\textbf{Weight} & \textbf{Spin} & \textbf{Semi-chiral ring element} \\
\hline
\endhead 

\hline
\multirow{1}{*}{$\frac{1}{3}$} & \multicolumn{1}{c|}{} & \multicolumn{1}{c|}{} \\
\hline
\multirow{1}{*}{$\frac{2}{3}$} & 0 & $\gamma \to \gamma$ \\
\hline
\multirow{1}{*}{1} & \multicolumn{1}{c|}{} & \multicolumn{1}{c|}{} \\
\hline
\multirow{1}{*}{$\frac{4}{3}$} & \multicolumn{1}{c|}{} & \multicolumn{1}{c|}{} \\
\hline
\multirow{1}{*}{$\frac{5}{3}$} & $\frac{1}{2}$ & $\partial_1 \gamma \to \pad \gamma$ \\
\hline
\multirow{1}{*}{2} & \multicolumn{1}{c|}{} & \multicolumn{1}{c|}{} \\
\hline
\multirow{1}{*}{$\frac{7}{3}$} & \multicolumn{1}{c|}{} & \multicolumn{1}{c|}{} \\
\hline
\multirow{1}{*}{$\frac{8}{3}$} & 1 & $\partial_1^2 \gamma \to \pad \partial_\beta \gamma$ \\
\hline
\multirow{1}{*}{3}
    & $\frac{1}{2}$ & $-\gamma \partial_1 \beta + 2 \partial_1 \gamma \beta \to S_\alpha$ \\
\hline
\multirow{1}{*}{$\frac{10}{3}$}
    & 1 & $-2\gamma \partial_1^2 \gamma + (\partial_1 \gamma)^2 \to \xi_{\alpha \beta}$ \\
\hline
\multirow{1}{*}{$\frac{11}{3}$}
    & $\frac{3}{2}$ & $\partial_1^3 \gamma \to \pad \partial_\beta \partial_\gamma \gamma$ \\
\hline
\multirow{2}{*}{4}
    & 1 & $-\gamma \partial_1^2 \beta + \partial_1 \gamma \partial_1 \beta + 2\partial_1^2 \gamma \beta \to \pad S_\beta$ \\
\cline{2-3}
    & 0 & $\partial_2 \gamma \partial_1 \beta - \partial_1 \gamma \partial_2 \beta \to \pad S^\alpha$ \\
\hline
\multirow{2}{*}{$\frac{13}{3}$}
    & $\frac{3}{2}$ & $-3\gamma \partial_1^3 \gamma + \partial_1 \gamma \partial_1^2 \gamma \to \pad \xi_{\beta \gamma}$ \\
\cline{2-3}
    & $\frac{1}{2}$ & $\partial_2 \gamma \partial_1^2 \gamma - \partial_1 \gamma \partial_1 \partial_2 \gamma \to \partial^\alpha \xi_{\alpha \beta}$ \\
\hline
\multirow{2}{*}{$\frac{14}{3}$}
    & 2 & $\partial_1^4 \gamma \to \pad \partial_\beta \partial_\gamma \partial_\delta \gamma$ \\
\cline{2-3}
    & 1 & $4\gamma^2 \partial_1^2 \beta + 6(\partial_1 \gamma)^2 \beta - 7\gamma \partial_1 \gamma \partial_1 \beta - 8\gamma \partial_1^2 \gamma \beta \to 9\pad \gamma S_\beta - 12\gamma \pad S_\beta$ \\
\hline
\multirow{2}{*}{5}
    & $\frac{3}{2}$ & $-\gamma \partial_1^3 \beta + 3\partial_1^2 \gamma \partial_1 \beta + 2\partial_1^3 \gamma \beta \to \pad \partial_\beta S_\gamma$ \\
\cline{2-3}
    & $\frac{1}{2}$ & $-\partial_2 \gamma \partial_1^2 \beta + \partial_1 \gamma \partial_1 \partial_2 \beta - \partial_1 \partial_2 \gamma \partial_1 \beta + \partial_1^2 \gamma \partial_2 \beta \to \pad \partial_\beta S^\beta$ \\
\hline
\clearpage
\multirow{3.5}{*}{$\frac{16}{3}$}
    \vspace{-0.5cm}
    & \multirow{2.1}{*}{2} & $-6\gamma \partial_1^4 \gamma + (\partial_1^2 \gamma)^2 \to \gamma \pad \partial_\beta \partial_\gamma \partial_\delta \gamma, \pad \partial_\beta \xi_{\gamma \delta}$  \\
\cline{3-3}
    &                     & $-4\gamma \partial_1^4 \gamma + \partial_1 \gamma \partial_1^3 \gamma \to \gamma \pad \partial_\beta \partial_\gamma \partial_\delta \gamma, \pad \partial_\beta \xi_{\gamma \delta}$ \\
\cline{2-3}
    & 1 & $\partial_2 \gamma \partial_1^3 \gamma - \partial_1 \gamma \partial_1^2 \partial_2 \gamma \to \pad \partial^\gamma \xi_{\gamma \beta}$ \\
\cline{2-3}
    & \multirow{2}{*}{0} & $\partial_2^2 \gamma \partial_1^2 \gamma - (\partial_1 \partial_2 \gamma)^2 \to \pad \partial_\beta \xi^{\alpha \beta}$ \\
    & & $-\gamma \partial_2 \beta \partial_1 \beta + 2 \partial_2 \gamma \beta \partial_1 \beta - 2 \partial_1 \gamma \beta \partial_2 \beta \to \Upsilon$ \\
\hline
\multirow{5.5}{*}{$\frac{17}{3}$}
    & $\frac{5}{2}$ & $\partial_1^5 \gamma \to \pad \partial_\beta \partial_\gamma \partial_\delta \partial_\epsilon \gamma$ \\
\cline{2-3}
    & \multirow{1.8}{*}{$\frac{3}{2}$} & $8\gamma^2 \partial_1^3 \beta + 7(\partial_1 \gamma)^2 \partial_1 \beta - 7\gamma \partial_1 \gamma \partial_1^2 \beta - 21\gamma \partial_1^2 \gamma \partial_1 \beta - 16\gamma \partial_1^3 \gamma \beta + 8\partial_1 \gamma \partial_1^2 \gamma \beta \to -24\gamma \pad \partial_\beta S_\gamma + 21\pad \gamma \partial_\beta S_\gamma - 9\pad \partial_\beta \gamma S_\gamma$ \\
\cline{2-3}
    & \multirow{3.1}{*}{$\frac{1}{2}$} & $-\gamma \partial_2 \gamma \partial_1^2 \beta + \gamma \partial_1 \gamma \partial_1 \partial_2 \beta + \gamma \partial_1 \partial_2 \gamma \partial_1 \beta - \gamma \partial_1^2 \gamma \partial_2 \beta + 4\partial_2 \gamma \partial_1^2 \gamma \beta - 4\partial_1 \gamma \partial_1 \partial_2 \gamma \beta \to -\gamma \pad \partial_\beta S^\beta + 6\pad \partial_\beta \gamma S^\beta$ \\
\cline{3-3}
    & & $-(\partial_1 \gamma)^2 \partial_2 \beta - \gamma \partial_2 \gamma \partial_1^2 \beta + \gamma \partial_1 \gamma \partial_1 \partial_2 \beta  - \gamma \partial_1 \partial_2 \gamma \partial_1 \beta + \gamma \partial_1^2\gamma \partial_2\beta + \partial_2\gamma\partial_1\gamma\partial_1\beta \to \pad \gamma \partial_\beta S^\beta - \gamma \pad \partial_\beta S^\beta$ \\
\hline
\multirow{4}{*}{6}
    & 2 & $-\gamma \partial_1^4 \beta - \partial_1 \gamma \partial_1^3 \beta + 3\partial_1^2 \gamma \partial_1^2 \beta + 5\partial_1^3 \gamma \partial_1 \beta + 2\partial_1^4 \gamma \beta \to \pad \partial_\beta \partial_\gamma S_\delta$ \\
\cline{2-3}
    & \multirow{1.8}{*}{1} & $\partial_2 \gamma \partial_1^3 \beta - \partial_1 \gamma \partial_1^2 \partial_2 \beta + 2\partial_1 \partial_2 \gamma \partial_1^2 \beta - 2\partial_1^2 \gamma \partial_1 \partial_2 \beta + \partial_1^2 \partial_2 \gamma \partial_1 \beta - \partial_1^3 \gamma \partial_2 \beta \to \pad \partial_\beta \partial_\gamma S^\gamma$ \\
\cline{2-3}
    & \multirow{1.8}{*}{0} & $-12\gamma (\partial_1 \partial_2 \gamma)^2 - \partial_2^2 \gamma (\partial_1 \gamma)^2 - \partial_1^2 \gamma (\partial_2 \gamma)^2 + 12\gamma \partial_2^2 \gamma \partial_1^2 \gamma + 2\partial_2 \gamma \partial_1 \gamma \partial_1 \partial_2 \gamma \to (S_\alpha \, S^\alpha)$ \\
\hline

\end{longtable}

\sectiono{The partition function and the index}
\label{partFapp}

We use the cohomology data of appendix \ref{semi-chiral ring app} (and its extension up to weight 10) to compute the partition function $Z \equiv \Tr{x^D z^{2 N_\beta + N_\gamma} p^{N_{\partial_1}} q^{N_{\partial_2}}}$ where $D$ is the dilatation Cartan, $N_\beta$ and $N_\gamma$ count the number of $\beta$ and $\gamma$ fields, and $N_{\partial_{1, 2}}$ count their derivatives, $F$ is the fermion number and the trace is over the cohomology. Note also that the $\mathbb{Z}_3$-fugacity $z$ is a third root of unity with $\beta$ having charge 2 and $\gamma$ having charge 1. The result is
\allowdisplaybreaks
\begin{align*}
        Z &=  1 +
	w^\frac{2}{3} z +
	w^\frac{5}{3} (p z + q z) +
	w^\frac{8}{3} (p^2 z + p q z + q^2 z) +
	w^3 (p + q)  +
 w^\frac{10}{3} (p^2 z^2 + p q z^2 + q^2 z^2) + \\&
 w^\frac{11}{3} (p^3 z + p^2 q z + p q^2 z + q^3 z) +
	w^4 (p^2 + 2 p q + q^2)  +
 w^\frac{13}{3} (p^3 z^2 + 2 p^2 q z^2 + 2 p q^2 z^2 + q^3 z^2) + \\&
	w^\frac{14}{3} (p^2 z + p^4 z + p q z + p^3 q z + q^2 z +  p^2 q^2 z +
    p q^3 z + q^4 z) +
w^5 (p^3 + 2 p^2 q + 2 p q^2 +
	q^3)  + \\&
     w^\frac{16}{3} (2 p^4 z^2 + p q z^2 + 3 p^3 q z^2 + 4 p^2 q^2 z^2 +
    3 p q^3 z^2 + 2 q^4 z^2) +
	w^\frac{17}{3} (p^3 z + p^5 z + 3 p^2 q z +  p^4 q z \\& + 3 p q^2 z +
    p^3 q^2 z + q^3 z + p^2 q^3 z + p q^4 z + q^5 z) +
	w^6 (p^4 + 2 p^3 q + 3 p^2 q^2 + 2 p q^3 +
    q^4)  + \\&
    w^\frac{19}{3} (2 p^5 z^2 + p^2 q z^2 + 4 p^4 q z^2 + p q^2 z^2 +
	5 p^3 q^2 z^2 +  5 p^2 q^3 z^2 + 4 p q^4 z^2 + 2 q^5 z^2) + \\&
	w^\frac{20}{3} (2 p^4 z + p^6 z + 5 p^3 q z + p^5 q z + 7 p^2 q^2 z +
    p^4 q^2 z + 5 p q^3 z + p^3 q^3 z + 2 q^4 z + p^2 q^4 z +
    p q^5 z + q^6 z) + \\&
       w^7(p^5 + p^2 q + 3 p^4 q + p q^2 + 4 p^3 q^2 +
    4 p^2 q^3 + 3 p q^4 + q^5) +
       w^\frac{22}{3} (3 p^6 z^2 + p^3 q z^2 + 5 p^5 q z^2 + 2 p^2 q^2 z^2 + \\&
    7 p^4 q^2 z^2 + p q^3 z^2 + 7 p^3 q^3 z^2 + 7 p^2 q^4 z^2 +
    5 p q^5 z^2 + 3 q^6 z^2) +
       w^\frac{23}{3} (3 p^5 z + p^7 z + 8 p^4 q z + p^6 q z + \\& 11 p^3 q^2 z +
    p^5 q^2 z + 11 p^2 q^3 z + p^4 q^3 z + 8 p q^4 z + p^3 q^4 z +
       3 q^5 z + p^2 q^5 z + p q^6 z + q^7 z) + \\&
        w^8 (2 p^6 + 2 p^3 q + 4 p^5 q +
       3 p^2 q^2 + 7 p^4 q^2 + 2 p q^3 +  7 p^3 q^3 + 7 p^2 q^4 +
    4 p q^5 + 2 q^6) +
       w^\frac{25}{3} (3 p^7 z^2 + \\& 2 p^4 q z^2 + 6 p^6 q z^2 + 4 p^3 q^2 z^2 +
       8 p^5 q^2 z^2 +  4 p^2 q^3 z^2 + 9 p^4 q^3 z^2 + 2 p q^4 z^2 +
       9 p^3 q^4 z^2 + 8 p^2 q^5 z^2 + \\& 6 p q^6 z^2 + 3 q^7 z^2) +
    w^\frac{26}{3} (4 p^6 z + p^8 z + 11 p^5 q z + p^7 q z + 16 p^4 q^2 z +
       p^6 q^2 z + 17 p^3 q^3 z + p^5 q^3 z + \\ &16 p^2 q^4 z + p^4 q^4 z +
       11 p q^5 z + p^3 q^5 z + 4 q^6 z + p^2 q^6 z + p q^7 z + q^8 z) +
  w^9 (p^5 + 2 p^7 + 4 p^4 q + \\ &6 p^6 q +
       7 p^3 q^2 + 10 p^5 q^2 + 7 p^2 q^3 + 12 p^4 q^3 + 4 p q^4 +
    12 p^3 q^4 + q^5 + 10 p^2 q^5 + 6 p q^6 + 2 q^7) + \\&
       w^\frac{28}{3} (p^6 z^2 + 4 p^8 z^2 + 4 p^5 q z^2 + 7 p^7 q z^2 +
    9 p^4 q^2 z^2 + 10 p^6 q^2 z^2 + 10 p^3 q^3 z^2 +
       11 p^5 q^3 z^2 + 9 p^2 q^4 z^2 + \\& 12 p^4 q^4 z^2 + 4 p q^5 z^2 +
    11 p^3 q^5 z^2 + q^6 z^2 + 10 p^2 q^6 z^2 + 7 p q^7 z^2 +
    4 q^8 z^2) +
       w^\frac{29}{3} (5 p^7 z + p^9 z + \\& 14 p^6 q z + p^8 q z + p^3 q^2 z +
    21 p^5 q^2 z + p^7 q^2 z + p^2 q^3 z + 24 p^4 q^3 z + p^6 q^3 z +
       24 p^3 q^4 z +  p^5 q^4 z + \\ & 21 p^2 q^5 z + p^4 q^5 z + 14 p q^6 z +
     p^3 q^6 z + 5 q^7 z + p^2 q^7 z + p q^8 z + q^9 z) +
    w^{10} (p^6 +
    3 p^8 + 7 p^5 q + \\& 8 p^7 q + 11 p^4 q^2 + 15 p^6 q^2 +
    13 p^3 q^3 + 18 p^5 q^3 + 11 p^2 q^4 + 21 p^4 q^4 + 7 p q^5 +
    18 p^3 q^5 + q^6 + 15 p^2 q^6 + \\& 8 p q^7 + 3 q^8) + \ldots
    \end{align*}
For the supersymmetric index $I \equiv \Tr{(-1)^F x^D z^{2 N_\beta + N_\gamma} p^{N_{\partial_1}} q^{N_{\partial_2}}}$, where $F$ is the fermion number, we obtain

\begin{align*}
    I &= 1 +
    w^\frac{2}{3} z +
    w^\frac{5}{3} (p z + q z) +
  w^\frac{8}{3} (p^2 z + p q z + q^2 z) +
  w^3(-p - q) +
  w^\frac{10}{3} (p^2 z^2 + p q z^2 + q^2 z^2) + \\&
  w^\frac{11}{3} (p^3 z + p^2 q z + p q^2 z + q^3 z) +
  w^4(-p^2 - 2 p q - q^2)  +
  w^\frac{13}{3} (p^3 z^2 + 2 p^2 q z^2 + 2 p q^2 z^2 + q^3 z^2) + \\&
  w^\frac{14}{3} (-p^2 z + p^4 z - p q z + p^3 q z - q^2 z + p^2 q^2 z +
    p q^3 z + q^4 z) +
    w^5 (-p^3 - 2 p^2 q - 2 p q^2 - q^3)+  \\&
     w^\frac{16}{3} (2 p^4 z^2 + p q z^2 + 3 p^3 q z^2 + 4 p^2 q^2 z^2 +
    3 p q^3 z^2 + 2 q^4 z^2) +
    w^\frac{17}{3} (-p^3 z + p^5 z - 3 p^2 q z + p^4 q z - \\& 3 p q^2 z +
    p^3 q^2 z - q^3 z + p^2 q^3 z + p q^4 z + q^5 z) +
    w^6 (-p^4 - 2 p^3 q -  p^2 q^2 - 2 p q^3 -
    q^4)+
    w^\frac{19}{3} (2 p^5 z^2 + \\& p^2 q z^2 + 4 p^4 q z^2 + p q^2 z^2 +
    5 p^3 q^2 z^2 + 5 p^2 q^3 z^2 + 4 p q^4 z^2 + 2 q^5 z^2) +
    w^\frac{20}{3} (-2 p^4 z + p^6 z - 5 p^3 q z + \\& p^5 q z - 7 p^2 q^2 z +
    p^4 q^2 z - 5 p q^3 z + p^3 q^3 z - 2 q^4 z + p^2 q^4 z +
    p q^5 z + q^6 z) +
     w^7 (-p^5 + p^2 q - p^4 q + \\& p q^2 - p q^4 -
    q^5)+
    w^\frac{22}{3} (3 p^6 z^2 + p^3 q z^2 + 5 p^5 q z^2 + 7 p^4 q^2 z^2 +
    p q^3 z^2 +7 p^3 q^3 z^2 + 7 p^2 q^4 z^2 + 5 p q^5 z^2 + \\&
    3 q^6 z^2) +
     w^\frac{23}{3} (-3 p^5 z + p^7 z - 8 p^4 q z + p^6 q z - 11 p^3 q^2 z +
    p^5 q^2 z - 11 p^2 q^3 z + p^4 q^3 z - 8 p q^4 z + \\ &p^3 q^4 z -
    3 q^5 z + p^2 q^5 z + p q^6 z + q^7 z) +
     w^8 (2 p^3 q + 3 p^2 q^2 + 3 p^4 q^2 + 2 p q^3 +
    3 p^3 q^3 + 3 p^2 q^4)+ \\&
    w^\frac{25}{3} (3 p^7 z^2 + 6 p^6 q z^2 - 2 p^3 q^2 z^2 + 8 p^5 q^2 z^2 -
    2 p^2 q^3 z^2 + 9 p^4 q^3 z^2 + 9 p^3 q^4 z^2 + 8 p^2 q^5 z^2 +
    6 p q^6 z^2 + \\ &3 q^7 z^2) +
    w^\frac{26}{3} (-4 p^6 z + p^8 z - 11 p^5 q z + p^7 q z - 16 p^4 q^2 z +
    p^6 q^2 z - 17 p^3 q^3 z + p^5 q^3 z -  16 p^2 q^4 z + \\& p^4 q^4 z -
    11 p q^5 z + p^3 q^5 z - 4 q^6 z +  p^2 q^6 z + p q^7 z + q^8 z) +
    w^9(p^5 + 4 p^4 q + 2 p^6 q +
    7 p^3 q^2 + 6 p^5 q^2 + \\ & 7 p^2 q^3 + 8 p^4 q^3 + 4 p q^4 +
    8 p^3 q^4 + q^5 + 6 p^2 q^5 + 2 p q^6)  +
     w^\frac{28}{3} (-p^6 z^2 + 4 p^8 z^2 - 2 p^5 q z^2 + 7 p^7 q z^2 - \\&
    7 p^4 q^2 z^2 + 10 p^6 q^2 z^2 - 8 p^3 q^3 z^2 + 11 p^5 q^3 z^2 -
    7 p^2 q^4 z^2 + 12 p^4 q^4 z^2 - 2 p q^5 z^2 + 11 p^3 q^5 z^2 -
    q^6 z^2 + \\& 10 p^2 q^6 z^2 + 7 p q^7 z^2 + 4 q^8 z^2)+
    w^\frac{29}{3} (-5 p^7 z + p^9 z - 14 p^6 q z + p^8 q z + p^3 q^2 z -
    21 p^5 q^2 z + p^7 q^2 z + \\& p^2 q^3 z - 24 p^4 q^3 z +
    p^6 q^3 z -
    24 p^3 q^4 z + p^5 q^4 z - 21 p^2 q^5 z + p^4 q^5 z - 14 p q^6 z +
     p^3 q^6 z - 5 q^7 z + \\& p^2 q^7 z + p q^8 z + q^9 z) +
    w^{10}(p^6 + p^8 +
    7 p^5 q + 4 p^7 q + 11 p^4 q^2 + 11 p^6 q^2 + 13 p^3 q^3 +
    14 p^5 q^3 + \\ & 11 p^2 q^4 + 17 p^4 q^4 + 7 p q^5 + 14 p^3 q^5 +
    q^6 + 11 p^2 q^6 + 4 p q^7 + q^8) + \ldots
   \end{align*}
which matches the expected expression
\begin{equation}
I = \prod_{n \geq 0} \prod_{m \geq 0} \frac{1-z^2 w^{n+m+\frac{4}{3}} p^n q^m}{1-z w^{n+m+\frac{2}{3}} p^n q^m}.
\end{equation}

\printbibliography
\end{document}